\documentclass{article}
\usepackage{authblk}

\usepackage[utf8]{inputenc}
\usepackage[T1]{fontenc}

\usepackage{amsmath}
\usepackage{graphicx}
\usepackage{xcolor}

\usepackage{hyperref}

\usepackage{listings}
\definecolor{mGreen}{rgb}{0,0.6,0}
\definecolor{mGray}{rgb}{0.45,0.45,0.45}
\definecolor{mPurple}{rgb}{0.58,0,0.82}
\definecolor{backgroundColour}{rgb}{0.97,0.97,0.97}
\definecolor{backgroundColourOP}{rgb}{0.97,0.97,0.97}

\lstdefinestyle{CStyle}{
    backgroundcolor=\color{backgroundColour},   
    commentstyle=\color{mGreen},
    keywordstyle=\color{magenta},
    numberstyle=\tiny\color{mGray},
    basicstyle=\ttfamily,
    breakatwhitespace=false,         
    breaklines=true,                 
    captionpos=b,                    
    keepspaces=true,                 
    numbers=left,                    
    numbersep=5pt,                  
    showspaces=false,                
    showstringspaces=false,
    showtabs=false,                  
    tabsize=2,
    language=C
}

\lstdefinestyle{ACSLStyle}{
    backgroundcolor=\color{backgroundColour},   
    commentstyle=\color{gray},
    keywordstyle=\color{magenta},
    numberstyle=\tiny\color{mGray},
    stringstyle=\color{mPurple},
    basicstyle=\ttfamily\scriptsize,
    breakatwhitespace=false,
    breaklines=true,
    captionpos=b,
    keepspaces=true,
    numbers=left,
    numbersep=5pt,
    showspaces=false,
    showstringspaces=false,
    showtabs=false,
    tabsize=2,
    language=C,
    deletecomment={[s]{/*}{*/}},
    moredelim=**[is][\ttfamily\scriptsize\color{mGreen}]{(@}{@)},
    moredelim=[is][\ttfamily\scriptsize\color{mGray}]{[@}{@]},
    morekeywords=[2]{assigns,requires,ensures,terminates},
    keywordstyle=[2]{\bfseries\color{mGreen}},
    belowskip=0pt
}

\usepackage{caption}
\usepackage{subcaption}

\usepackage{tikz}
\usetikzlibrary{positioning, automata}
\usetikzlibrary{shapes}
\usetikzlibrary{arrows}
\usetikzlibrary{fit}

\usepackage{lmodern}

\newcommand{\tool}[1]{\textsf{#1}}
\newcommand{\autodeduct}{\tool{AutoDeduct}}
\newcommand{\saida}{\tool{Saida}}
\newcommand{\isp}{\tool{ISP}}
\newcommand{\tricera}{\tool{TriCera}}

\newcommand{\framac}{\tool{Frama-C}}
\newcommand{\wpre}{\tool{WP}}
\newcommand{\eva}{\tool{Eva}}

\begin{document}

\pagestyle{plain}

\title{\autodeduct{}: A Tool for Automated Deductive Verification of C Code}

\author[1]{Jesper Amilon}
\author[1]{Dilian Gurov}
\author[2]{Christian Lidstr\"{o}m}
\author[1,3]{Mattias Nyberg}
\author[1,3]{Gustav Ung}
\author[3]{Ola Wingbrant}

\affil[1]{KTH Royal Institute of Technology, Stockholm, Sweden}
\affil[2]{Fondazione Bruno Kessler, Trento, Italy}
\affil[3]{Scania AB, S\"odert\"alje, Sweden}



\maketitle

\begin{abstract}
Deductive verification has become a mature paradigm for the verification of industrial software. Applying deductive verification, however, requires that every function in the code base is annotated with a function contract specifying its behaviour. This introduces a large overhead of manual work. To address this challenge, we introduce the \autodeduct{} toolchain, built on top of the \framac{} framework. It implements a combination of techniques to \emph{automatically infer contracts} for functions in C~programs, in the syntax of ACSL, the specification language of \framac{}. Contract inference in \autodeduct{} is implemented as two plugins for \framac{}, each inferring different types of annotations. We assume that programs have an entry-point function already equipped with a contract, which is used in conjunction with the program source code to infer contracts for the helper functions, so that the entry-point contract can be verified. The current release of \autodeduct{} is the first public prototype, which we evaluate on an example adapted from industrial software.
\end{abstract}

\section{Introduction}
Over the past 6 years, we have used the \framac{} framework for the deductive verification of industrial automotive software, together with the heavy-vehicles manufacturer Scania~\cite{nybe-et-al-2018,ung-et-al-24,skantz-21-msc,manjikian-23-msc}. A key insight from our experience is that deductive verification imposes a huge manual overhead to annotate the code with specifications, and that  
this is a major obstacle for deductive verification to become a practically viable option for the industry.
To overcome this obstacle, we have 
worked towards developing 
techniques that \emph{automatically infer} ACSL specifications. In this paper, we report on the first prototype of our toolchain \autodeduct{}, which automates the deductive verification process of C programs
through a combination of inference techniques.

When working with programs in languages such as C, each file typically has only a few C~functions serving as \emph{entry-points}, but they are implemented through numerous \emph{helper} functions. To apply deductive verification, every function is expected to be annotated with a contract. Often, one also needs to add further annotations for the verification to succeed, such as annotations specifying memory safety or which memory location a function is allowed to access.
In our toolchain, we implement several techniques for \emph{property-guided} contract inference, to automatically generate specifications that otherwise must be provided manually.
As we illustrate in the evaluation (\autoref{sec:evaluation}), \autodeduct{} is well-suited for deductive verification of C code in embedded systems, where requirements typically specify a module, and not individual C functions.



\paragraph{Related work.}

The \emph{weakest pre-condition calculus}, pioneered by Dijkstra~\cite{Dijkstra:1975:GCN:360933.360975}, was an early method for automatic specification inference: given a post-condition, the weakest pre-condition sufficient to establish the post-condition can be computed through syntactic predicate transformations defined over program statements. 
As presented by Moy~\cite{DBLP:conf/vmcai/Moy08}, pre-conditions that are not necessarily the weakest can be computed by a combination of abstract interpretation and quantifier elimination.
Seghir et al. presented an approach based on \emph{Counterexample-Guided Abstraction Refinement} (CEGAR) to inferring pre-conditions~\cite{DBLP:conf/esop/SeghirK13,DBLP:conf/aplas/SeghirS14}.

Applying the dual notion of \emph{strongest post-conditions}~\cite{Dijkstra:1975:GCN:360933.360975} (or \emph{function summaries}) 
is another approach to specification inference. Strongest post-conditions are computed from a program and a pre-condition, and characterise the final states that result from executions from states satisfying the pre-condition.
Strongest post-conditions can be computed through symbolic execution~\cite{DBLP:books/daglib/p/GordonC10} for programs without unbounded loops. Sinleton et al. proposed an algorithm for transforming exponentially large post-conditions resulting from strongest post-condition computation into more concise forms~\cite{DBLP:journals/corr/abs-1905-06847}.

\emph{Maximal specification inference} generalises weakest pre-condition inference by considering specifications for multiple functions simultaneously. Albarghouthi et al. present a counter-example guided approach, which is based on multi-abduction~\cite{DBLP:conf/popl/AlbarghouthiDG16}.

\paragraph{Contribution.} 
This paper presents the first public prototype release of the \autodeduct{} toolchain (\autoref{sec:autodeduct}), using a novel combination of techniques to automatically infer contracts for helper functions in the context of an entry-point contract, and off-the-shelf technology for verification of C programs.
It is built on top of the \framac{}~\cite{kirchner15-frama-c,frama-c-manual} framework, and infers function contracts for usage in deductive verification in \framac{}. The contract inference in the toolchain consists of two separate \framac{} plugins, performing complementary forms of contract inference.
The paper also presents an initial evaluation on a C program adapted from industrial software (\autoref{sec:evaluation}).

\section{Background}
\label{sec:background}

\framac{}~\cite{frama-c-manual} is an open-source modular platform for static analysis of C programs. It features plugins such as \wpre{}~\cite{wpplugin}, used for deductive verification based on the weakest pre-condition calculus, and \eva{}~\cite{evaplugin}, used for showing absence of runtime errors by abstract interpretation. \framac{} also allows users to write new plug-ins to extend its capabilities.




\begin{figure}[t]
\centering
\hspace{1em}
\begin{minipage}{0.45\linewidth}
\begin{lstlisting}[style=ACSLStyle]
int read_input() {
  return in;
}
void write_output(int x) {
  out = x;
}

int saturate(int x, int lim){ 
  if (x > lim) return lim;
  else return x;
}
\end{lstlisting}
\end{minipage}
\hspace{1em}
\begin{minipage}{0.45\linewidth}
\begin{lstlisting}[style=ACSLStyle,firstnumber=12]
(@/*@ requires in >= 0;
    ensures
      in <= 10 ==> out == in*in
      && in > 10 ==> out == 100;   
*/@)
void main() {
  int tmp = read_input();
  tmp = saturate(tmp, 10);
  tmp = tmp * tmp;
  write_output(tmp);
}
\end{lstlisting}
\end{minipage}
\caption{Example program with a main function and three helper functions, where only the main function is given a contract.}
\label{fig:simple_example_code}
\end{figure}

The specifications to be verified deductively with \framac{} are provided to the program as annotations written in the ANSI C Specification Language (ACSL)~\cite{acsl}. In particular, ACSL offers \emph{function contracts}, which are used to specify the functional behaviour of a given C~function. An example of an ACSL function contract can be seen in \autoref{fig:simple_example_code}. The contract is primarily expressed through the \lstinline{requires} and \lstinline{ensures} clauses. The former clause asserts that a condition must hold directly before the function call, and has to be shown to hold at each call site. The latter clause asserts that a condition must hold directly upon return. Variables occurring in this clause can refer either to their value in the post-state of the function, or in the pre-state if used with the \lstinline{\old} predicate. In addition to these conditions, a frame condition may also be added, specifying which memory locations may be modified by the function, and is expressed through the \lstinline{assigns} keyword.



\section{The \autodeduct{} Toolchain}
\label{sec:autodeduct}

\autodeduct{} is a toolchain for automated deductive verification of C programs. The intended use case is a C~module, for the \emph{entry-point} function of which an ACSL contract has been provided. \autodeduct{} will infer contracts for the helper functions, and then verify the program using \wpre{}.
The verification with \wpre{} ensures that the entry-point function satisfies the provided contract, and that every helper function satisfies its inferred contract. Thus, the necessary trust base for the verification engineer does not extend beyond the mature \wpre{} plugin of \framac{}. Aside from verifying the program, the toolchain produces as an artefact a fully annotated program. In the domain of contract inference tools, we consider \autodeduct{} as a toolchain for \emph{property-guided} contract inference, meaning that it infers contracts that are strong enough for verifying the current program, but they may not generalise to other contexts.


We classify annotations within ACSL contracts as either \emph{functional} or \emph{auxiliary}~\cite{lids-thesis-24}. The functional annotations directly concern the behaviour of the function, typically by relating input values to output values of program variables. 
Auxiliary annotations are those that do not directly specify the functional behaviour, but may be needed for the verification to succeed, e.g., to specify memory safety. 
Previously, we have worked on these inference paradigms separately, but with the \autodeduct{} tool, we bring them together under one roof.  


\autodeduct{} is implemented in the form of two separate \framac{} plugins, one for functional annotations and the other one for auxiliary annotations.
The architecture of the \autodeduct{} toolchain is shown in \autoref{fig:tool-architecture}.
\autodeduct{} is available as a Docker image on its Github page\footnote{\url{https://github.com/rse-verification/auto-deduct-toolchain}}, or in the accompanying artifact. 

We shall use \autoref{fig:simple_example_code} as a running example to explain the auxiliary and functional inference components of \autodeduct{}, and illustrate its usage. The example is a C~program that structurally resembles code from the embedded systems domain. The entry-point of the program reads the input, performs some calculations, and writes the output. Specifically, it sets the output to the maximum value of 100 and the square of the input. The main function uses three helper functions, \lstinline$read_input$, \lstinline$saturate$, and \lstinline$write_output$. The ACSL contract for the main function specifies the intended behaviour of the program. By running \autodeduct{} on the program, we infer contracts for the helper functions, and verify the program using \wpre{}. The inferred contracts for \lstinline$saturate$ and \lstinline$read_input$ are shown in \autoref{fig:simple_example_inferred_contracts}. Note that we have formatted the contracts for readability. 


\begin{figure*}[t]
\centering
\scalebox{1}{\newif\ifdrawtriceratools
\drawtriceratoolsfalse
\newif\ifdrawpreprocessor
\drawpreprocessorfalse
\newif\ifdrawlegend
\drawlegendtrue

\pgfdeclarelayer{bg}    
\pgfdeclarelayer{bbg}    
\pgfsetlayers{bbg,bg,main}  

\begin{tikzpicture}[%
  default/.style={%
    draw,
    rounded corners,
    inner sep = 0.4em,
    minimum height = 2em,
    align = center},
]
\scriptsize
\node (code) [default, sharp corners] {C module};
\node (input) [below = 3em of code, inner sep=0pt, outer sep = 0pt] {};
\node (reqs) [below = 1em of input, default, sharp corners] {ACSL\\specification};
\draw[-] (code) to (input);
\draw[-] (reqs) to (input);

\ifdrawpreprocessor
\node (preproc) [right = 5em of input, default, fill=blue!20] {Pre-\\processor};
\node (saida) [right = 5em of preproc, default, fill=blue!20, align = center] {Functional\\ contract\\ inference};
\draw[->] (input) to (preproc);
\draw[->] (preproc) to (saida);
\else
\node (saida) [right = 10em of input, default, fill=blue!20, align = center] {Functional\\ contract\\ inference};
\draw[->] (input) to (saida);
\fi

\node (isp) [right = 2em of saida, default, fill=blue!20] {Auxilliary\\ contract\\ inference};

\node (eva) [below = 1em of isp, default, fill=blue!20, align=center] {Eva};
\draw[<->] (isp) to (eva);
\node (tricera) [above = 1.5em of saida, default, fill=blue!20] {Tricera};

\ifdrawtriceratools
\node (eldarica) [right = 1em of tricera, default, fill=yellow!30] {Eldarica};
\node (princess) [right = 1em of eldarica, default, fill=yellow!30] {Princess};
\fi

\draw[->] (saida) to (isp);
\draw[<->] (saida) to (tricera);
\ifdrawtriceratools
\draw[<->] (tricera) to (eldarica);
\draw[<->] (eldarica) to (princess);
\fi

\node (wp) [right = 5em of isp, default, fill=orange!30] {WP};
\node (output) [right = 5em of wp, inner sep=0em, outer sep=0em] {};
\node (success) [above = 1em of output, default, sharp corners,text=green!90!black] {Verification\\succeeded};
\node (fail) [below = 1em of output, default, sharp corners,text=red!90!black] {Verification\\failed};

\draw[->] (isp) to node[above, align=center] {\tiny{Annotated}\\ \tiny{C file}} (wp);
\draw[->] (wp) to (success);
\draw[->] (wp) to (fail);

\begin{pgfonlayer}{bg}
\node (frama-c) [fit={([shift={(0pt,5pt)}]isp.north) ([shift={(-42pt,0pt)}]saida.west)  ([shift={(5pt,0pt)}]wp.east) (isp) (eva) (saida) (wp)},draw=gray!90] {};
\node[below right=0.022em and 0.022em, inner sep=4pt] at (frama-c.north west) {Frama-C};
\end{pgfonlayer}

\ifdrawtriceratools
\begin{pgfonlayer}{bg}
\node (tricera-tools) [fit={(tricera) (eldarica) (princess)},draw=gray!90, inner sep=5pt, fill=yellow!10] {};
\end{pgfonlayer}
\fi

\ifdrawtriceratools
\node (autodeduct-north) [above = 0pt of tricera-tools, inner sep = 0pt] {};
\else
\node (autodeduct-north) [above = 0pt of tricera, inner sep = 0pt] {};
\fi
\ifdrawpreprocessor
\node (autodeduct-west) [left = 2pt of preproc, inner sep = 0pt] {};
\else
\node (autodeduct-west) [left = 46pt of saida, inner sep = 0pt] {};
\fi

\begin{pgfonlayer}{bbg}
\node (autodeduct) [fit={([shift={(0pt,15pt)}]isp.north)  (autodeduct-west) (autodeduct-north) (frama-c)},draw=gray!90] {};
\node[below right=0.022em and 0.022em, inner sep=4pt] at (autodeduct.north west) {\textbf{AutoDeduct}};
\end{pgfonlayer}

\ifdrawlegend
\node (legend-anchor) [below = 10 pt of autodeduct] {};
\node (legend-autodeduct) [left = 80pt of legend-anchor, inner sep = 5pt, draw, rounded corners, fill=blue!20, label=right:Contract inference] {};
\node (legend-frama-c) [right = 105pt of legend-autodeduct, inner sep = 5pt, draw, rounded corners, fill=orange!30, label=right:Verification] {};
\fi

\end{tikzpicture}}
\caption{Architecture of \autodeduct{}}
\label{fig:tool-architecture}
\end{figure*}





  
  


\begin{figure}[!htb]
    \centering
\begin{subfigure}{\textwidth}
\begin{lstlisting}[style=ACSLStyle]
(@/*@ requires 0 <= x <= 2147483647;  [@// Auxiliary@]
    requires lim == 10 && x == in;  [@// Functional@]
    ensures  [@// Functional@]
      0 <= \result <= 10 && out == \old(out) && 
      in == \old(in) && \old(lim) == 10 && \old(x) == \old(in) &&
      (\result == \old(in) ==> (10 >= \old(in) >= 0)) &&
      (\result == 10 ==> \old(in) >= 10) &&
      (\result == \old(in) || \result == 10);
    assigns \nothing; */ [@// Auxiliary@] @)
int saturate(int x, int lim);

(@/*@ requires 0 <= x <= 100 && \valid(&out); [@// Auxiliary@] 
    requires [@// Functional@]
      (in * in == x && 0 <= in <= 10) || (10 * 10 == x && in >= 11);
    ensures 0 <= out <= 100; [@// Auxiliary@]
    ensures [@// Functional@]
  ((in * in == \old(x) && 0 <= in <= 10) || (10 * 10 == \old(x) && in >= 11)) 
        && in == \old(in) && \old(x) == out;
    assigns out; */ [@// Auxiliary@] @)
void write_output(int x);
\end{lstlisting}
\end{subfigure}

    \caption{Inferred contracts for the program \lstinline!write_output! and \lstinline!saturate!}
    \label{fig:simple_example_inferred_contracts}
\end{figure}

\paragraph{Functional contract inference.}
The functional contract inference component of \autodeduct{} is a \framac{} plugin~\cite{amilon-et-al-24} (previously referred to as the \saida{} plugin), which assumes
%
that the entry-point is annotated with an ACSL contract. It then uses as backend the Horn-clause based model checker \tricera{} for inference of functional aspects of the helper functions contracts.  
\tricera{} encodes the program as a set of Horn clauses, where each called function is associated with a pair of predicates representing the pre-condition and post-condition. An interpretation for these predicates represents sufficient conditions for the program to be \emph{correct}. Based on this solution, contracts for the helper functions can be extracted. For further technical details on the interplay between \tricera{} and \autodeduct{}, we refer to our earlier work~\cite{alshnakat-et-al-20,amilon-et-al-24}.

After running the functional inference component, we obtain a contract for every helper function. For example, for the \lstinline!saturate! function, the key parts of the functional aspect of the contract are the two conjuncts on lines 6 and~7. The contract also illustrates that inferred contracts can be highly context-specific; for example, the contract for \lstinline!saturate! requires \lstinline!lim == 10!, as this is the value of this parameter at the only function call.


\paragraph{Auxiliary contract inference.}

The auxiliary contract inference component of \autodeduct{} is implemented as another \framac{} plugin~\cite{skantz-21-msc,manjikian-23-msc} (referred to as the \isp{} plugin in previous work). For the inferred contracts in \autoref{fig:simple_example_inferred_contracts}, the auxiliary aspects include specifying ranges for variable values, the validity of pointers, and memory locations that are assigned to. All these annotations are derived from an abstract interpretation of the program, using the 
\framac{} plugin \eva{} as the backend. 
%
The specification for pointer validity ensures that all pointers used in the program are valid, as shown in the contract for \lstinline$write_output$. 
The \lstinline{assigns} clauses stem from the analysis by \eva{}; for example, the contract for \lstinline$write_output$ now specifies that only the variable \lstinline!out! may be assigned to. 

\paragraph{Running the toolchain.}
The \autodeduct{} toolchain is intended to be used mainly as a command-line tool. It assumes as input a C program, where the entry-point function is equipped with an ACSL contract, and can be run by invoking a script: \lstinline!./autodeduct filename.c!. The user may also run each step separately as follows. The functional contract inference component is run by \lstinline{./autodeduct -func filename.c}, which runs the functional inference and outputs the file \lstinline$tmp_inferred_source_merged.c$. Next, the auxiliary contract inference is invoked with \lstinline$./autodeduct -aux tmp_inferred_source_merged.c$, which outputs the now fully annotated file \lstinline!annotated.c!. The fully annotated program can be verified with \wpre{} using \lstinline{./autodeduct -wp annotated.c}.

\paragraph{Limitations.}
While we support a relatively large subset of C and ACSL, there are currently some limitations in the toolchain. In general, the auxiliary inference component supports a larger subset than the functional inference component. Notable features not supported for functional inference are floating-point arithmetic and stack pointers. C~features supported by neither functional nor auxiliary inference include pointer arithmetic, nested pointers, and local static variables. \autodeduct{} also does not support the general use of logic functions in ACSL contracts. While the toolchain supports programs with loops, we have yet to implement the automated inference of loop invariants.

\section{Evaluation}
\label{sec:evaluation}

We evaluate \autodeduct{} on a case study consisting of a C module of $ca.\ 1400$ LOC, taken from the automotive industry. We have previously verified the module using \framac{}~\cite{nybe-et-al-2018} by 
manually annotating it,
in an effort that took 2 person-months. We have also shown previously that the auxiliary annotation component of \autodeduct{} can infer almost all of the needed auxiliary annotations~\cite{manjikian-23-msc}. 

Due to the current limitations in the toolchain, we consider a simplified version (which can be found in the accompanying artifact) of the module and 5 of its requirements.
In the simplified module, which consists of $123$ LOC, we have removed the parts not relevant for the specified requirements and, e.g., rewritten certain pointer constructs.
It has an entry-point function annotated with a function contract, and 5~helper functions without contracts. 
The contract specifies 5~requirements on the module, formulated by engineers at Scania, each one represented by a separate \lstinline{ensures} clause in the contract, and an \lstinline{assigns} clause that specifies the variables that the module may write to. 

When applying \autodeduct{} on the simplified module, the verification succeeds in roughly 23~s,
%
of which 18~s are spent on functional inference, 2~s on auxiliary, and 3~s on verification with \wpre{}. Moreover, the toolchain produces as output a fully annotated version of the module, with inferred contracts for all the helper functions.
The inferred contracts contain the necessary \lstinline!requires!, \lstinline!ensures!, and \lstinline{assigns} clauses.
We did not compare \autodeduct{} against other tools, since we are not aware of any existing toolchain for deductive verification that can automatically verify such C~programs using contract inference.

\section{Conclusion}\label{sec:conclusion}
We have presented the \autodeduct{} toolchain for automated deductive verification of C code, based on \framac{} and a combination of techniques for contract inference.
We believe our toolchain to be the first one to offer such a high degree of automation.
The case study illustrates our vision and shows that {\autodeduct{} can automatically verify industrial C~code specified with industrial requirements. 
Future work will foremost focus on covering all features of~C and ACSL needed to automatically verify the full version of the module used here and other modules taken from the automotive industry.
%
%
We also plan to enhance the user feedback in the case of failed verification results. 

\textbf{}

\bibliographystyle{splncs04}
\bibliography{refs}

\begin{thebibliography}{10}
\providecommand{\url}[1]{\texttt{#1}}
\providecommand{\urlprefix}{URL }
\providecommand{\doi}[1]{https://doi.org/#1}

\bibitem{DBLP:conf/popl/AlbarghouthiDG16}
Albarghouthi, A., Dillig, I., Gurfinkel, A.: Maximal specification synthesis. In: Proceedings of the 43rd Annual {ACM} {SIGPLAN-SIGACT} Symposium on Principles of Programming Languages, {POPL} 2016, St. Petersburg, FL, USA, January 20 - 22, 2016. pp. 789--801 (2016). \doi{10.1145/2837614.2837628}

\bibitem{alshnakat-et-al-20}
Alshnakat, A., Gurov, D., Lidstr{\"o}m, C., R{\"u}mmer, P.: Constraint-Based Contract Inference for Deductive Verification, pp. 149--176. Springer International Publishing, Cham (2020). \doi{10.1007/978-3-030-64354-6\_6}

\bibitem{amilon-et-al-24}
Amilon, J., Esen, Z., Gurov, D., Lidstr{\"o}m, C., R{\"u}mmer, P.: An Exercise in Mind Reading: Automatic Contract Inference for {Frama-C}, pp. 553--582. Springer International Publishing, Cham (2024). \doi{https://doi.org/10.1007/978-3-031-55608-1\_13}

\bibitem{acsl}
Baudin, P., Filli\^{a}tre, J.C., March\'{e}, C., Monate, B., Moy, Y., Prevosto, V.: {ACSL}: {ANSI/ISO C} Specification Language (2024), \url{http://frama-c.com/download/acsl.pdf}

\bibitem{wpplugin}
Baudin, P., Bobot, F., Correnson, L., Dargaye, Z., Blanchard, A.: {WP} Plug-in Manual. CEA LIST (2024), \url{https://frama-c.com/download/frama-c-wp-manual.pdf}

\bibitem{evaplugin}
Bühler, D., Cuoq, P., Yakobowski, B.: Eva - The Evolved Value Analysis plug-in. CEA LIST (2024), \url{http://frama-c.com/download/frama-c-eva-manual.pdf}

\bibitem{frama-c-manual}
Correnson, L., Cuoq, P., Kirchner, F., Maroneze, A., Prevosto, V., Puccetti, A., Signoles, J., Yakobowski, B.: Frama-{C} User Manual. CEA LIST, Inria, \url{http://frama-c.com/download/frama-c-user-manual.pdf}

\bibitem{Dijkstra:1975:GCN:360933.360975}
Dijkstra, E.W.: Guarded commands, nondeterminacy and formal derivation of programs. Commun. ACM  \textbf{18}(8),  453--457 (Aug 1975). \doi{10.1145/360933.360975}

\bibitem{DBLP:books/daglib/p/GordonC10}
Gordon, M., Collavizza, H.: Forward with {Hoare}. In: Reflections on the Work of C. A. R. Hoare., pp. 101--121. Springer (2010). \doi{10.1007/978-1-84882-912-1\_5}

\bibitem{kirchner15-frama-c}
Kirchner, F., Kosmatov, N., Prevosto, V., Signoles, J., Yakobowski, B.: {Frama-C: A Software Analysis Perspective}. Formal Aspects of Computing  (2015). \doi{10.1007/978-3-642-33826-7\_16}

\bibitem{lids-thesis-24}
Lidstr{\"o}m, C.: Automated Deductive Verification of Safety-Critical Embedded Software. Ph.D. thesis, KTH, Theoretical Computer Science, TCS (2024), qC 20240223

\bibitem{manjikian-23-msc}
Manjikian, H.: Improving the Synthesis of Annotations for Partially Automated Deductive Verification. Master's thesis, KTH Royal Institute of Technology (2021)

\bibitem{DBLP:conf/vmcai/Moy08}
Moy, Y.: Sufficient preconditions for modular assertion checking. In: Verification, Model Checking, and Abstract Interpretation, 9th International Conference, {VMCAI} 2008, San Francisco, USA, January 7-9, 2008, Proceedings. pp. 188--202 (2008). \doi{10.1007/978-3-540-78163-9\_18}

\bibitem{nybe-et-al-2018}
Nyberg, M., Gurov, D., Lidstr\"{o}m, C., Rasmusson, A., Westman, J.: Formal verification in automotive industry: Enablers and obstacles. In: Leveraging Applications of Formal Methods, Verification and Validation. Industrial Practice: 8th International Symposium, ISoLA 2018, Proceedings, Part IV. p. 139–158. Springer-Verlag, Berlin, Heidelberg (2018). \doi{10.1007/978-3-030-03427-6\_14}

\bibitem{DBLP:conf/esop/SeghirK13}
Seghir, M.N., Kroening, D.: Counterexample-guided precondition inference. In: Programming Languages and Systems - 22nd European Symposium on Programming, {ESOP} 2013, Held as Part of the European Joint Conferences on Theory and Practice of Software, {ETAPS} 2013, Rome, Italy, March 16-24, 2013. Proceedings. pp. 451--471 (2013). \doi{10.1007/978-3-642-37036-6\_25}

\bibitem{DBLP:conf/aplas/SeghirS14}
Seghir, M.N., Schrammel, P.: Necessary and sufficient preconditions via eager abstraction. In: Programming Languages and Systems - 12th Asian Symposium, {APLAS} 2014, Singapore, November 17-19, 2014, Proceedings. pp. 236--254 (2014). \doi{10.1007/978-3-319-12736-1\_13}

\bibitem{DBLP:journals/corr/abs-1905-06847}
Singleton, J.L., Leavens, G.T., Rajan, H., Cok, D.R.: Inferring concise specifications of {APIs}. CoRR  \textbf{abs/1905.06847} (2019), \url{http://arxiv.org/abs/1905.06847}

\bibitem{skantz-21-msc}
Skantz, D.: Synthesis of Annotations for Partially Automated Deductive Verification. Master's thesis, KTH Royal Institute of Technology (2021)

\bibitem{ung-et-al-24}
Ung, G., Amilon, J., Gurov, D., Lidström, C., Nyberg, M., Palmskog, K.: Post-hoc formal verification of automotive software with informal requirements: An experience report. In: 2024 IEEE 32nd International Requirements Engineering Conference (RE). pp. 287--298 (2024). \doi{10.1109/RE59067.2024.00035}

\end{thebibliography}

\end{document}